\documentclass{aa}
\usepackage{graphicx}
\usepackage{rotating}
\usepackage{txfonts}
\usepackage{epsf}
\usepackage{natbib}

\def\new#1{{ #1}}

\def\del#1{{\it \small #1}}
\def\del#1{{}}
\def\oii{{\sc{[O\,ii]}}}

\def\GOODS{{\sc Goods}\ }
\def\FLAMES-GIRAFFE{{\sc Flames-Giraffe}}
\def\GIRAFFE{{\sc Giraffe}}
\def\GALFIT{{\sc Galfit}}

\begin{document}
\title{A  surviving  disk from a galaxy collision at $z=0.4$}
\author{
       Y. Yang\inst{1}                     
  \and F. Hammer\inst{1}                   
  \and H. Flores\inst{1}                   
  \and M. Puech\inst{2,1}                  
  \and M. Rodrigues\inst{1}                
}
\offprints{yanbin.yang@obspm.fr}
\authorrunning{Y. Yang et al.}
\titlerunning{A temporarily surviving disk}
\institute{
GEPI, Observatoire de Paris, CNRS, Universit\'e Paris Diderot; 5 Place Jules Janssen, Meudon, France
\and
ESO, Karl-Schwarzschild-Strasse 2, D-85748 Garching bei M\"unchen, Germany
}
\date{Received ...... ; accepted ...... }

\abstract
{Spiral galaxies dominate the local galaxy population.
Disks are known to be fragile with respect to collisions.
Thus it is  worthwhile to probe under which conditions
a disk can possibly survive  such interactions.}
{We present a detailed morpho-kinematics study of a
    massive galaxy with two nuclei,
  \,\object{J033210.76--274234.6}, \,at $z=0.4$.}
{The morphological analysis reveals that the object
  consists of two bulges and a massive disk, as well as a faint
    blue ring.  Combining the kinematics with morphology we propose a
      near-center collision model to interpret the object.}
{\new{We find that the massive disk is likely to have
  survived the collision of galaxies with an
  initial mass ratio of $\sim$\,4\,:\,1.
  The N-body/Smoothed Particle Hydrodynamics (SPH) simulations show
  that the collision possibly is a single-shot polar collision
  with a very small pericentric distance of $\sim$\,1\,kpc,
  and that the remnant of the main galaxy will be dominated by a disk.
  The results support the disk survival hypothesis. }}
{\new{The survival of the disk is related to the polar collision
with an extremely small pericentric distance.
With the help of N-body/SPH simulations
we find the probability of disk survival is quite large
regardless whether the two galaxies merge or not.}
}


\maketitle


\section{Introduction}
In the local universe, most of the intermediate-mass
galaxies are spiral galaxies \citep{2004AJ....127.2511N}
which have experienced a violent merging stage over the last 8
Gyrs \citep{2000MNRAS.311..565L, 2004ApJ...601L.123B,2008ApJ...681.1089R}.
How the local spirals have been formed is still a matter of
debate. \citet{2005A&A...430..115H} propose a scenario of
\,``disk-rebuilding'' based on the remarkable coincidence of
the evolution of the merger rate, morphology and fraction of
actively star-forming galaxies. In such a scenario, a
significant fraction of galaxies have experienced their last
major merger during the last 8 Gyrs, then, by accretion of
gas and debris, the disks were formed around the merger
remnants.

As a prediction of hierarchical models of structure formation
\citep{1978MNRAS.183..341W}, galaxy-galaxy mergers or interactions
have been an essential recipe to interpret the formation and evolution
of galaxies \citep{1992ARA&A..30..705B, 2000MNRAS.312..859S}.  Earlier
numerical simulations predicted that the remnants of major mergers of
disk galaxies (with mass ratio smaller than $\sim$\,4\,:\,1) are likely to
be elliptical galaxies \citep[e.g.,][]{1988ApJ...331..699B,
  1992ARA&A..30..705B, 1992ApJ...400..460H}, implying that the disks
of the progenitors are easily destroyed in major mergers.  The
simulations by \citet{2002MNRAS.333..481B} suggest that gas disks may
be rebuilt after major mergers from the orbital angular
momentum. Similar results are shown by recent simulations of gas-rich
encounters \citep{2005ApJ...622L...9S, 2006ApJ...645..986R,
  2008arXiv0806.1739H}.  While minor mergers are less violent than
major ones, the effect of minor mergers in re-shaping progenitors is
significant, even with large mass ratios of 10:1
\citep{1992ApJ...389....5T}.  By the accretion of satellite galaxies,
a thin disk will become thicker due to the dynamical heating by the
satellite in-fall. Although the disks of spiral galaxies may survive
minor mergers, their properties, such as the shape, the
star-formation, the bulge-to-disk ratio, are evolving towards the
early type spirals \citep{1996ApJ...460..121W}.

One interesting galaxy interaction is head-on collision which
may cause the formation of the ring in disk galaxies, so called
``collisional'' ring galaxies \citep{1976ApJ...209..382L}.  In this
case, the satellite galaxy impacts close to the center of the host
galaxy, and passes through the disk. Then a ring will be formed and
propagate in the disk outwards due to the density wave. Simulations
show that in most of the collisional ring galaxies, disks are severely
affected \citep{1996FCPh...16..111A,1997MNRAS.286..284A}.
\new{Therefore it is worthwhile to explore whether a disk will survive
the collision, and under what conditions,
such as mass ratio and orbital parameters.}

We present a morphology and kinematics study of J033210.76--274234.6
which is possibly an
indication of such a surviving disk. This paper is organized as
follows: in Sect.~\ref{secobs} we describe the data we have obtained
for the object; in Sect.~\ref{secanalysis}, the analysis of morphology
is presented. In Sect.~\ref{secmodel}, we develop a dynamical model
\new{together with N-body/SPH simulations in order to}
explain the observations. We discuss the results and give our
conclusions in Sect.~\ref{secdiscuss}.
Throughout the paper, we adopt the
Concordance cosmological parameters of
$H_0\!=\!70$\,km\,s$^{-1}$\,Mpc$^{-1}$, $\Omega_M\!=\!0.3$ and
$\Omega_\Lambda\!=\!0.7$.

\begin{figure*}[!t]
\centering
\begin{tabular}{cc}
 \includegraphics[height=6.5cm]{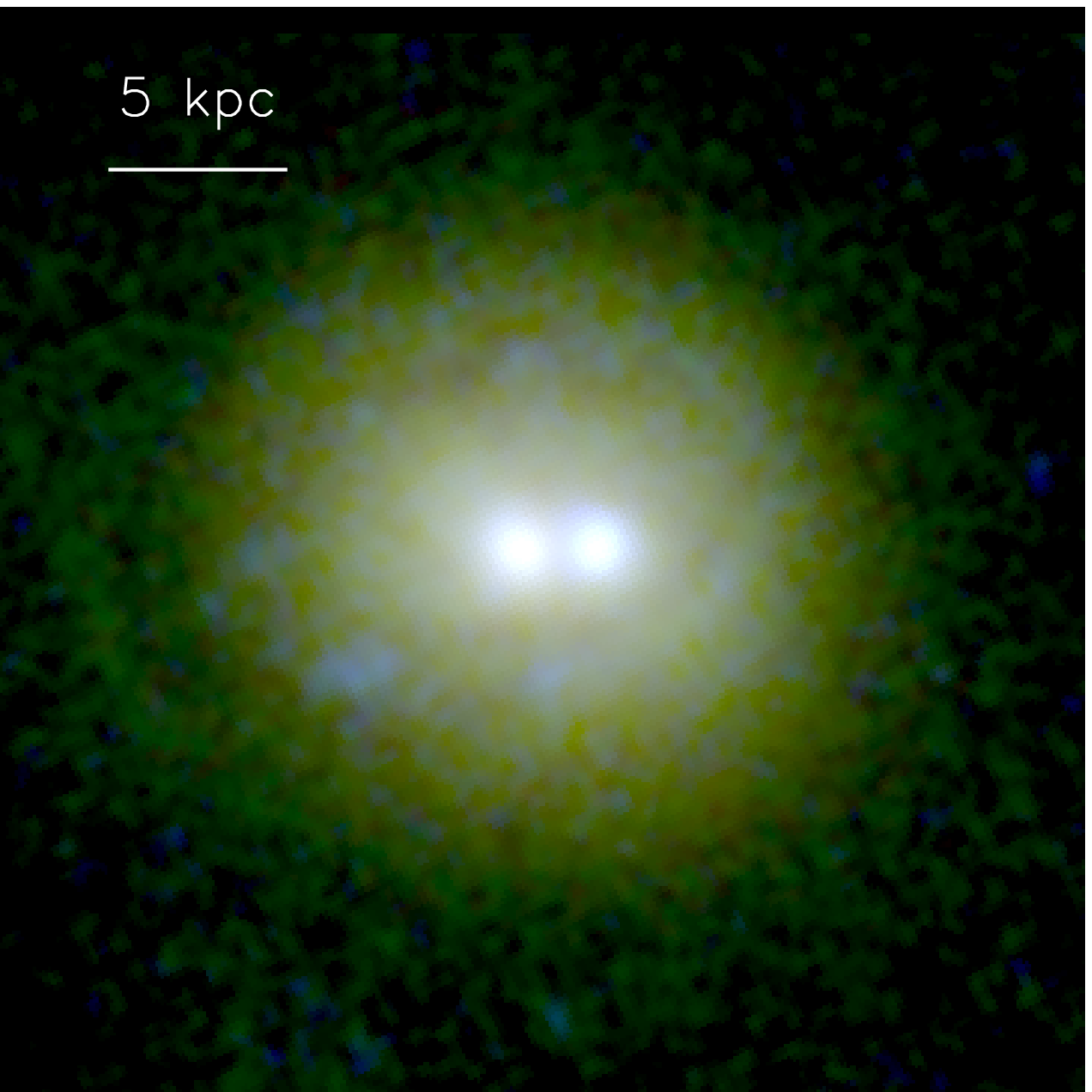}\hspace{3mm}\includegraphics[height=6.5cm]{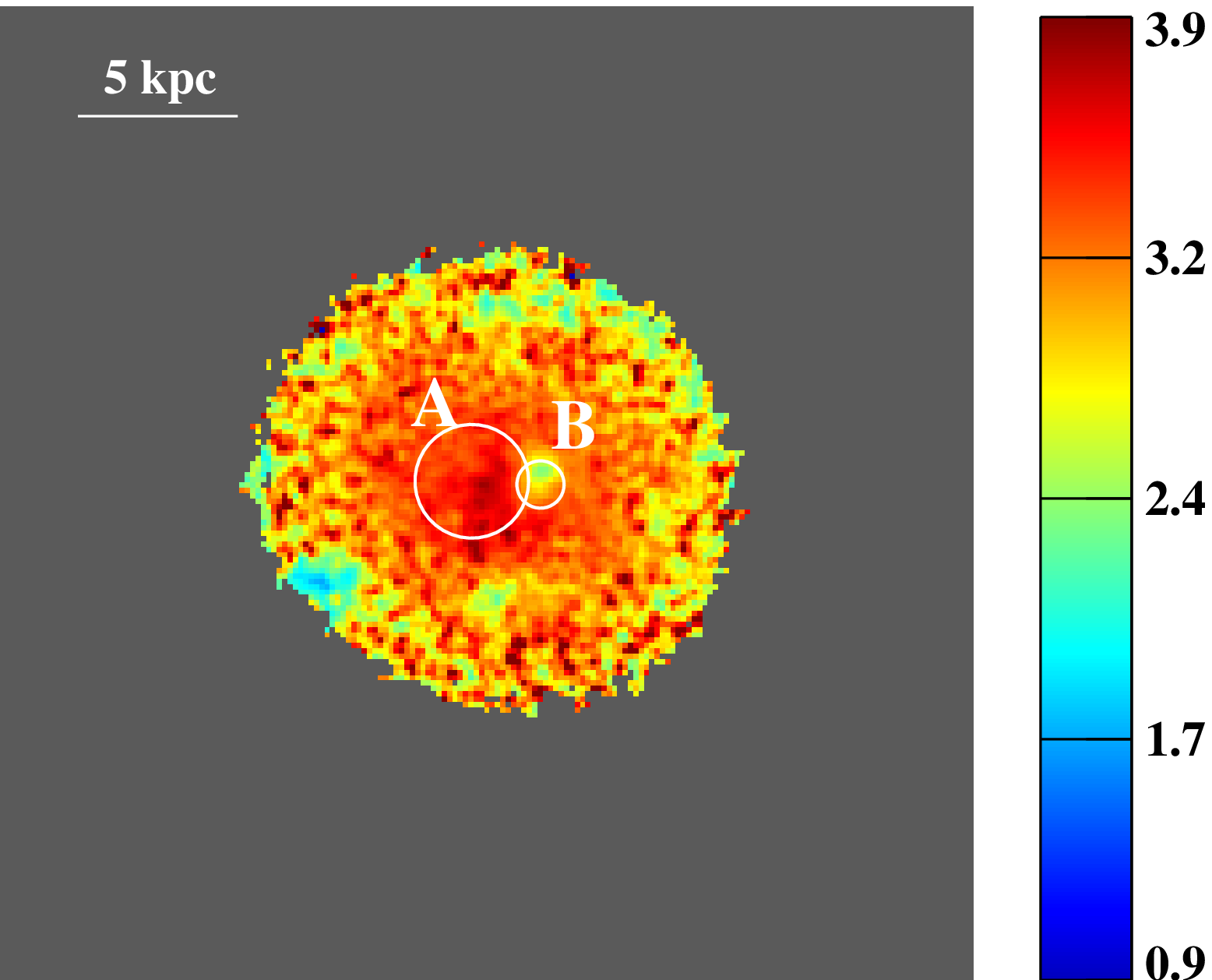}
\end{tabular}
\caption{{\it Left:}
the three-color image composed from the B, V, z  bands. \new{The image has a size of $30$\,kpc}.
{\it Right:} B$-$z color map  of the object.
\new{The regions with signal-to-noise ratio greater than 3 are shown.
The circles A, B indicate core-A and core-B, respectively.
The radii of the two circles are equivalent to
the effective radii of the two cores.
These two panels have identical scale and orientation. }}
\label{figcolors}
\end{figure*}

\section{Data}
\label{secobs}

The object
\object{J033210.76--274234.6} ($z$\,$=$\,$0.41686$)
is located in the \GOODS field
\citep[The Great Observatories Origins Deep Survey,][]{2004ApJ...600L..93G}.
The public data of \GOODS are available for the detailed
studies of its colors and morphology. We chose the data release 1.0
from {\it G{\tiny OODS} Cutout
  Service}\footnote{http://archive.stsci.edu/eidol.php} and obtained
the images in 4 bands: F435W, F606W, F775W, F850LP, namely B, V, i,
z. The drizzled HST/ACS images have a pixel scale of ${\rm
  0.03''/pixel}$, corresponding to 0.165 kpc/pixel at the target
redshift.  In Table~\ref{tbgalfit} we list the photometric properties
of the object.

With the \FLAMES-GIRAFFE multi-object integral-field spectrograph at
VLT \citep[see][]{2006A&A...455..107F}, the spatially-resolved kinematics of this object has been recovered by observing
the [\ion{O}{ii}] doublet emission \citep{2008A&A...477..789Y}. The
object was early selected into the sample of emission line galaxies
\citep{2008A&A...477..789Y}. From the high quality FORS2 spectroscopic observation at VLT
\citep{Rodrigues2008}, it is discovered to be a rather quiescent
galaxy with weak [\ion{O}{\sc ii}] emission (at $\sim$ 2\,\AA). Due to
the brightness of the object (V(AB)=19.9) we are able to recover its 
kinematics with \GIRAFFE\ at VLT.

\section{Analysis}
\label{secanalysis}
\subsection{General properties}
In Fig.~\ref{figcolors}, we show a three-color image and a B$-$z color
map of the object.  The color of the three-color image is slightly
enhanced for better recognizing the structures.  Two bright nuclei can
be easily recognized.  We have plotted two circles in the B$-$z color map to
indicate the core positions and the sizes at effective radius that
are derived from the morphology analysis (see
Sect.~\ref{secmorpholgy}). The left core (hereafter core-A), has a
color as red as the elliptical galaxies at redshift 0.4
\citep[see Figure~8 of][]{2008A&A...484..159N}. The right core
(hereafter core-B) is slightly bluer than core-A, and its color is
similar to that of an S0 galaxy.  In addition, we detect several blue
clumps surrounding the object center. Since it resembles a ring, we
refer to it as the ``blue ring'' hereafter. It is detected in B-band
which corresponds to U-band at rest frame. This suggests that
star formation has been triggered in the ring.

\begin{table}[!t]
\caption{Properties of the target \object{J033210.76--274234.6}.}
\label{tbgalfit}
{\tiny
\begin{tabular}{ccccccc}
\hline \hline
~&&&&&\\*[-2mm]
\multicolumn{4}{l}{\bf {\small Photometric properties:}} \\
  $M_B$(AB) &  $M_J$(AB) &   $R_{\rm half}$$^{*}$ & $M_{\rm stellar}^{\rm total}$ & SFR$_{\rm IR}$ & SFR$_{\rm 2800}$ & $A_V$ \\*[2pt]
            &            & (kpc) & ($10^{10} M_\odot$)  &     ($M_\odot$/yr) &   ($M_\odot$/yr) & \\
   $-$21.78 & $-$23.70   & 4.36 &$27.5 $ & 5.27  & 3.24 & 0.2 \\ \hline
\end{tabular}}
\vglue 1.5mm
\begin{tabular}{cccccc}
\multicolumn{5}{l}{\bf Morphological decomposition for z-band}  \\*[1pt]
& Magnitude & $R_{\rm eff}$$^{\rm a}$\,(kpc)&  F$_{i}$$^{\rm b}$ & ~~~B$-$z$^{\rm c}$ \\*[3pt] \hline\\*[-9pt]
\multicolumn{1}{l|}{core-A} & 20.03$\pm$0.01  &  1.78$\pm$0.02& 0.27$\pm$0.01 & ~~~3.53 \\
\multicolumn{1}{l|}{core-B}& 20.85$\pm$0.01  &  0.74$\pm$0.01& 0.13$\pm$0.01 & ~~~3.15  \\
\multicolumn{1}{l|}{Disk \hphantom{Ring}}     & 19.14$\pm$0.01&  5.39$\pm$0.01& 0.60$\pm$0.01 & ~~~2.94
\\ \hline\\*[-6pt]
\end{tabular}
\\
\begin{tabular}{ccccc}
& Center$^{\rm d}$ & P.A.$^{\rm e}$&  Inc.\,(b\,/\,a)$^{\rm f}$ & ~$M_{\rm stellar}$$^{\rm g}$  \\
\multicolumn{1}{l}{} & (kpc) & ($^\circ$)   &  &  ($10^{10} M_\odot$)   \\ \hline\\*[-9pt]
\multicolumn{1}{l|}{core-A} &  [0.00,   0.00] &~~~69  & 0.75   & ~~~\,3.57   \\
\multicolumn{1}{l|}{core-B} &  [2.15,$-$0.10]     &~~~65  & 0.82   & ~~~\,7.43   \\
\multicolumn{1}{l|}{Disk}     &[0.67,   0.03]     &~~~65  & 26$^\circ$   & 16.5   \\
\multicolumn{1}{l|}{Blue Ring}&[0.02,   0.12]  &$-$78 & 52$^\circ$ & --- \\ \hline\\*[-7pt]
\end{tabular}
\\
\begin{tabular}{ccccc}
\multicolumn{5}{l}{\bf Physical scales} \\
\multicolumn{2}{l}{The distance of the two cores} & 2.15 kpc &~~~~~~~~~~~~ &~~~~~~~~~~~~ \\
\multicolumn{2}{l}{The radius of the blue ring} & 7.18 kpc \\
\hline \hline
\end{tabular}
\vglue 2mm
$^{\rm *}$ The half-light radius.
$^{\rm a}$ The effective radius of the S\'{e}rsic profile.
$^{\rm b}$ The light fraction of each model component with respect to the total model flux.
$^{\rm c}$ The B$-$z color of each component, calculated based on the morphological decomposition.
$^{\rm d}$ The center of each component, measured with respect to the center of the disk.
$^{\rm e}$ The position angle of each component.
$^{\rm f}$ For the disk and the ring we give the inclination in degrees; for the bulges we give the axis ratios.
$^{\rm g}$ The stellar mass of each component (see Sect.~\ref{secmodel} for details).
\end{table}

\begin{figure*}[!ht]
\centering
\includegraphics[width=6.5cm]{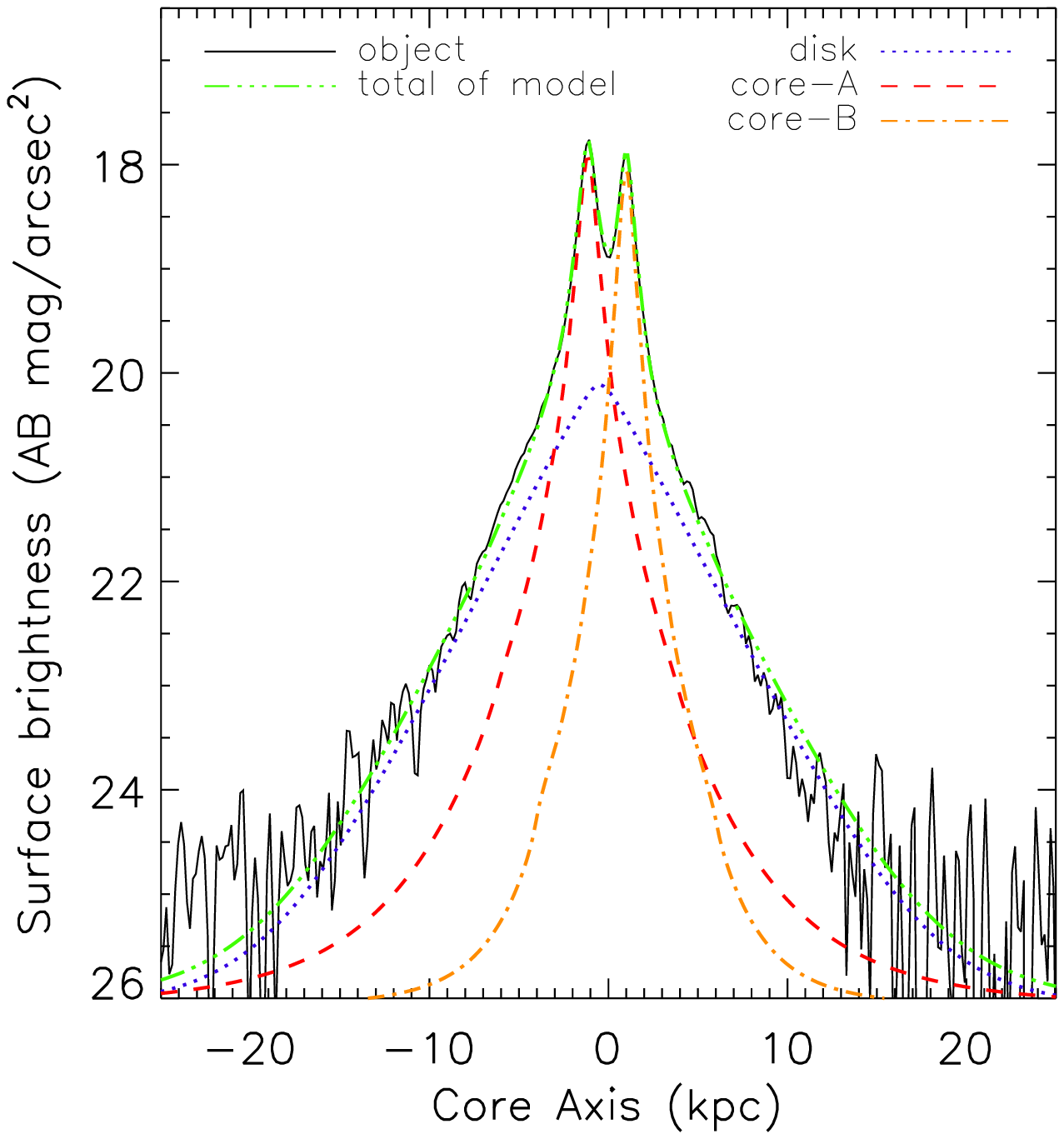}\includegraphics[width=6.5cm]{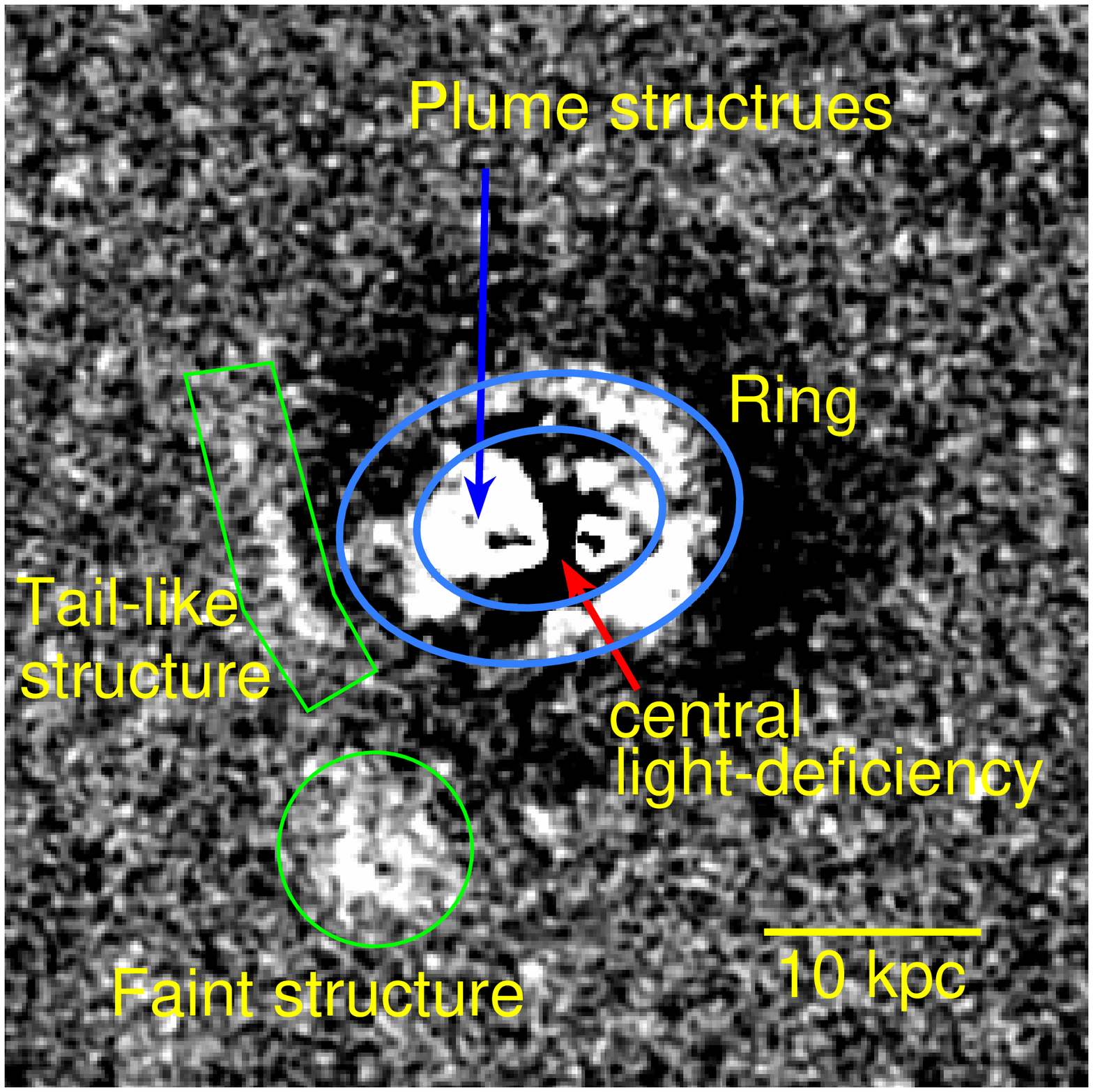}
\caption{{\it Left:} light profile decomposition along the core-axis of
  z-band. {\it Right:} the V-band residual image (slightly smoothed) of morphology
  decomposition.
  \new{The blue annulus indicates the ring region.
  The other two green regions indicate the tail-like structure and the faint structure, respectively.
  The plume structures and the central light-deficiency are marked by arrows as well. }}
\label{majoraxis}
\end{figure*}

\subsection{Morphological analysis}
\label{secmorpholgy}
\GALFIT\ \citep{2002AJ....124..266P} is used to perform a
two-dimensional profile decomposition in order to study the intrinsic
luminosity distribution. The technical details about the decomposition
have been discussed in \citet{2008A&A...484..159N} and
\citet{2007A&A...469..483R}. We chose the S\'{e}rsic profile to model
each component. The key parameter is the S\'{e}rsic index $n$, which
gives the classical bulge profile of the de Vaucouleurs law when
$n=4$, and the exponential disk when $n=1$. 

We first investigated the light profile of z-band.
We initially set a model of two S\'{e}rsic profiles that are centered
at the two nuclei respectively. {\GALFIT} gives a best fit with
$\chi^2/\nu$ of 1.88.  The significant residuals imply the existence
of an additional component.  We then added the third component of
S\'{e}rsic to the model.  We get an optimized solution of
$\chi^2/\nu$\,=\,1.26 with $n=5.67$, \,$4.61$, \,$0.48$ for
core-A, core-B and the third component, respectively. The two nuclei
are bulges since their $n$ approximate to 4. The third component
resembles the disk shape according to its lower $n$.  We have tried
a four-component model which gave no improvement. 
\new{Alternatively, we have tried to bind the disk center to any of the cores,
but no reasonable solutions were found. }
It is known that the S\'{e}rsic index is coupling to the effective radius
\citep{2002AJ....124..266P, 2001MNRAS.326..869T}. Thus we fixed the
index $n$ to the classical values: $n=4$ for bulges and $n=1$ for
the disk-like component. We obtained a reasonable fit with a
$\chi^2/\nu$ of 1.34. Although $\chi^2$ of the free-$n$ fit is smaller
than that of the fixed-$n$ fit, it does not significantly improve the
modeling. So we adopt the fixed-$n$ model hereafter.
We have performed
fixed-$n$ fitting to the other three bands (i.e., B, V, i). The
findings are similar to what we get from z-band. As a result, we
have successfully recovered the main components of the object, i.e.,
the two bulges and the large disk centered between the two.  The
parameters of each component in z-band and the error estimation by
{\GALFIT} are listed in Table~\ref{tbgalfit}.

The left panel of Fig.~\ref{majoraxis} illustrates the results of the
morphological decomposition. We show the light profile along the
core-axis which is defined to be the line across the centers of the
two bulges. The profile of each seeing-convolved model component as
well as the sum of them are plotted separately in Fig.~\ref{majoraxis}
with different symbols.
The light profile at $|r|>5$ kpc is
close to a straight line, which is strongly indicative of the existence
and the significance of the disk. The center of the disk is found to
be located between the two nuclei (close to core-A), and no
significant structures, e.g., arms, are found in the disk.  \new{Note that
in the residual maps of all the bands, we find several structures
(see V-band for example, the right panel in Fig.~\ref{majoraxis}),
including a shell-like structure following the blue ring,
plume-like structures that are associated with core-A, 
a tail-like structure and a faint structures.  
These residuals indicate a recent galactic interaction in the object.}

\begin{figure*}[!t]
\centering
\includegraphics[width=12cm]{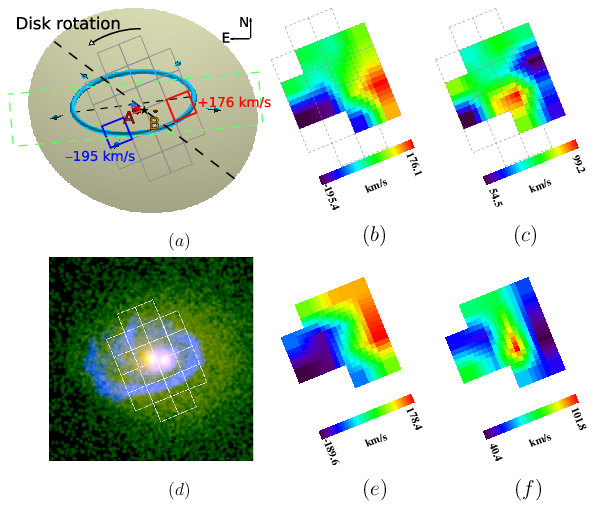}
\caption{
  {Panel ($a$)}, a 3D sketch of our target based on the
  morphological decomposition. The disk is represented by a
  gray plane. The two bulges are indicated by different
  colors, red for core-A and orange for core-B. The blue
  ring is shown as well in blue.
  Two black dashed lines indicate the major axes of
  the ring and the disk, respectively. The black star between
  the two bulges represents the center of the disk. The blue
  star above the core-A indicates the center of the ring.
  The rotated grid in gray lines indicates the \GIRAFFE\ IFU
  elements (see panels ($b$) and ($c$)). The red and the blue
  boxes highlight the maximal and the minimal velocities
  observed by \GIRAFFE\ IFU. The green
  dashed box indicates the position and the coverage of the
  FORS2 slit. Panels ($b$) and ($c$), the spatially-resolved kinematics, \new{i.e.,}
  the velocity field and the velocity dispersion, respectively.
  The dashed-line grids indicate the spatial
  resolution elements of IFU. The kinematics maps
  have been rotated to be aligned with the object
  orientation. \new{Panel ($d$) shows the simulated image which is constructed
  by the projected star density (in yellow-green) with the gas superposed in blue color.}
  Panels ($e$) and ($f$) show
  the simulated gas kinematics:  VF- and $\sigma$-maps  
  (see Sect.~\ref{secmodel} for more details).
  }
\label{figmodel}
\end{figure*}

\section{Dynamical model}
\label{secmodel}
\subsection{Analysis}
The detection of a ring suggests that
an almost central collision took place between two progenitors.
The morphological analysis
shows a very strong indication that the disk of main progenitor was
relatively little affected by this interaction,
as weak perturbations are detected in the residuals.
Below, with the help of \new{N-body/SPH
\citep[e.g.,][]{1992ARA&A..30..543M}}
simulations, we construct a model for this collision
in order to test whether the main progenitor disk may have survived
the interaction.

With the spatially-resolved kinematics from \GIRAFFE, the slit kinematics
from FORS2, and the morphological decomposition, we are able to
develop a dynamical model to interpret the interaction occurring
in the object. The mass of each component is crucial to construct the
model. We have obtained the stellar mass of the system from
\citet{2008A&A...484..173P} and listed it in Table~\ref{tbgalfit}.  We
estimate the stellar mass of each component according to their \new{z-band} light
fraction (see Table~\ref{tbgalfit}), under the assumption that the
luminosity in z-band (corresponding to R-band at rest frame) is
approximately proportional to the stellar mass.

The panel ($a$) of Fig.~\ref{figmodel} shows a 3D
sketch of the object based on the morphological parameters
that are derived from the decomposition. The disk is nearly
face-on with an inclination of $26^\circ$. The two bulges
have a projected distance of 2.15\,kpc. The blue ring with
an inclination of $52^\circ$ is also shown in the figure.
The ring has a radius of 7.18\,kpc.
The two plumes (Fig.~\ref{majoraxis}) may be
the tidal relics of the core collision, because they have a
red color similar to core-A which is marginally affected by
the global low extinction $A_V$\,=\,0.2 (see
Table~\ref{tbgalfit}). Moreover, we find that the center of
core-B in B-band is slightly shifted from the centers in the
other bands by 0.06$''$ ($\sim 0.3\,$kpc). This shift may
account for the asymmetric color distribution of core-B (see
B$-$z map Fig.~\ref{figcolors}).

\new{What are the progenitors?
The detection of the giant disk is robust.
This disk does not seem to be severely affected by the collision.
We refer to the galaxy associated with this disk as the main galaxy.
We have noticed that the B$-$z color of the disk is 2.94
which is in agreement with the color of S0 galaxies at this redshift
\citep[see Figure~8 of][]{2008A&A...484..159N}.
From its location, brightness and color, core-A is likely to be
the bulge of the main galaxy.
If we take core-B as the bulge of the main galaxy,
resulting in a bulge-to-total light ratio of B/T\,$\sim$\,0.17,
then we would expect a bluer disk such as Sbc type galaxies.
The inclination of the blue ring
is quite high compared to that of the detected disk.
This suggests that the blue ring is probably not lying in the plane of the disk.
It implies that another disk existed before the collision,
and this disk had been disrupted during the collision and formed the ring.
From the morphology decomposition, we have estimated 
the bulge mass of the two progenitors:
for the main galaxy, the bulge takes 27\% of the total mass of the system,
and for the intruder, the bulge takes 13\%.
It is difficult to estimate the exact mass ratios of the progenitors,
since the intruder's disk had been disrupted.
However, we can estimate the limits.
We have observed the star-forming activities, i.e., the blue ring
which suggests a significant fraction of gas preexisted
in the intruder disk.
First, let us assume the intruder is an Sa galaxy
with B/T of 0.3
which is a typical value for Sa galaxies.
With this assumption, we arrive at a mass ratio of $\sim$1\,:\,1
between the progenitors.
This seems not the case that we have observed.
Otherwise we would have seen two rings with similar strength.
Thus, the intruder could be an S0 galaxy with B/T ranging from 0.5 to 0.8.
We assume a mean B/T value of 0.65 for the intruder,
which gives a mass ratio of 4\,:\,1 for the progenitors.
Even if we take $\rm B/T=0.8$, the mass ratio only increases to 5:1 which could be
the upper limit of mass ratio.
We adopt a mass ratio of 4\,:\,1 in the following discussion,
resembling a major merger as usually mentioned.
}

\begin{figure}[!t]
\centering
\includegraphics[width=8cm]{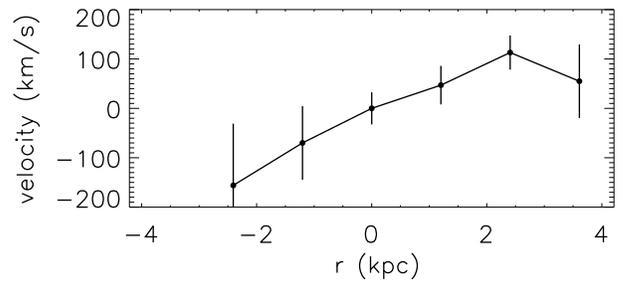}
\caption{The velocity curve of the object,
which is derived from the FORS2 slit observation.
The coverage of slit is shown in the panel ($a$)
of Fig.~\ref{figmodel} in a green dashed-line box.}
\label{figvc}
\end{figure}

\begin{figure*}[!t]
\centering
\includegraphics[width=6cm]{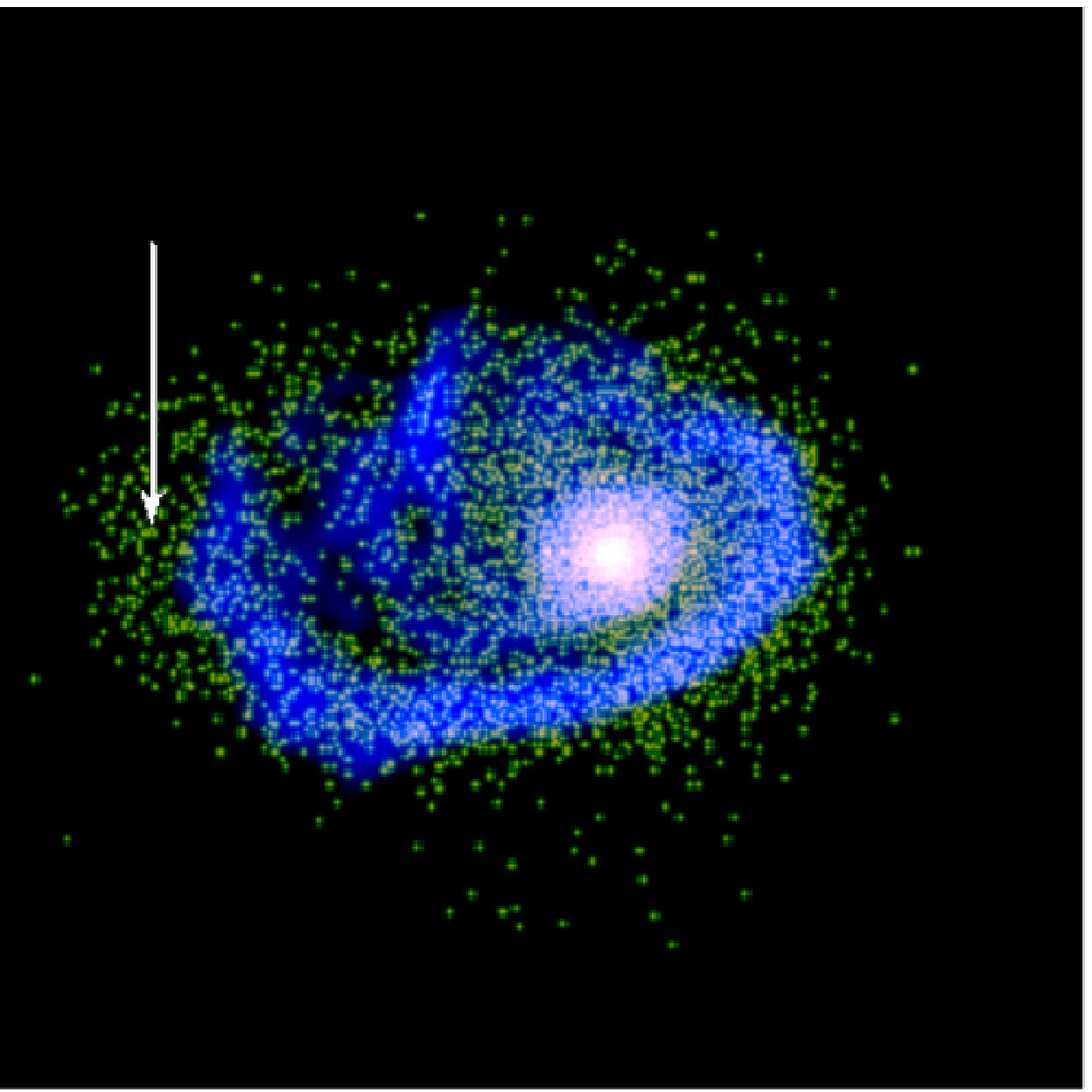}\hspace{1cm}\includegraphics[width=6cm]{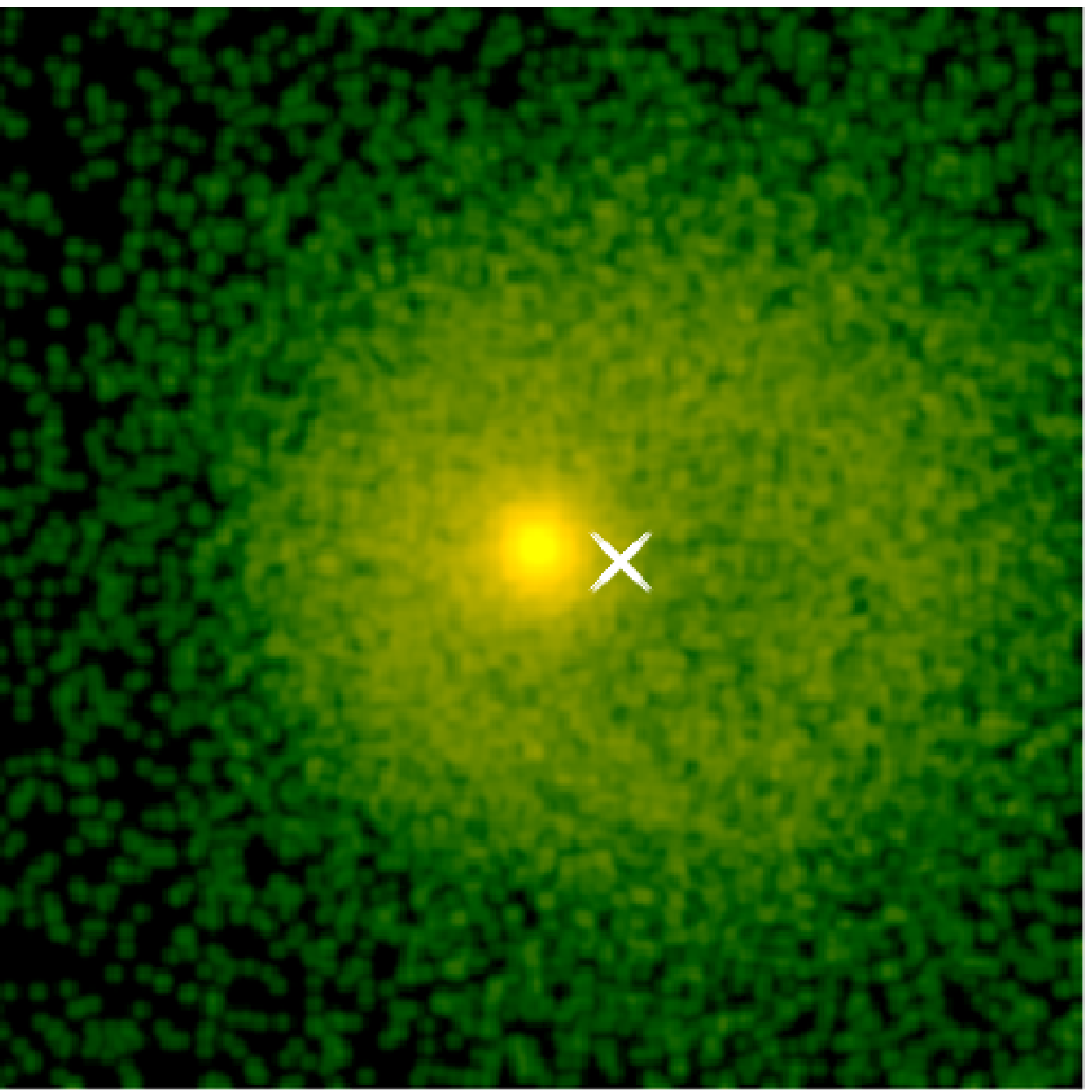} \\
 ($a$) \hspace{6cm} ($b$)
\caption{Panel ($a$), the simulated image for the intruder only, 
with gas superposed in blue color.
The arrow indicates the star distribution
which is possibly corresponding to the observed tail-like structure.
Panel ($b$), the simulated image for the main galaxy only.
A weak but broad ring can be recognized. 
Due to the superposition of the bulge, 
the left part of the ring shape is hidden. 
The cross indicates the center of the disk.
See Sect.~\ref{secsimulation} for more details.}
\label{figfallmain}
\end{figure*}

We have obtained FORS2 slit observations of the object
from \citet{Rodrigues2008}, see Fig.~\ref{figmodel} for the
slit coverage. Using the absorption lines
\ion{Ca}{{\sc ii} H, K}, we are able to recover a velocity curve covering
the central 7.2 kpc region (see Fig.~\ref{figvc}), which
suggests a disk dominated by rotation. With the maximal
$\Delta V$ of 269 km\,s$^{-1}$, the rotation velocity is
estimated to be 360\,km\,s$^{-1}$ after correcting for disk
inclination. \new{With the mass ratio of 4:1, we can derive
the stellar mass of the main galaxy
to be 2.2\,$\times$\,$10^{11}$\,$M_\odot$ (80\% of the total)
We find that this galaxy is
well in agreement with the stellar mass Tully-Fisher relation
by \citet{2007A&A...466...83P} and with the $R_d$-$V_{\rm
  flat}$ relation by \citet{2007ApJ...662..322H}.}

The right panel of Fig.~\ref{figmodel} shows the \GIRAFFE\ 2D
velocity field (VF) and velocity dispersion ($\sigma$) map.
Since the maps are derived from the [O\,{\sc ii}] emission,
the kinematics tracks the large-scale motion of the gas. The
dynamical axis, which is defined as the line connecting the
minimum and maximal velocity, is misaligned with the major
axis of the ring. This suggests that the motion of gas is
not a simple rotation pattern, and that another motion is
superposed on it.
The velocity gradient follows the ring. The above facts are
easily explained by the superposition of a disk rotation and
an expanding velocity field related to the ring.
We estimate the expanding velocity of
the ring to be $180\,\pm\,70$\,km\,s$^{-1}$ by taking into
account the geometric projection and several
\GIRAFFE\ measurements around the ring. This suggests the
elapsed time after the impact to be 39\,Myrs by taking the ring
radius of 7.18\,kpc.

\subsection{N-body/SPH simulation}
\label{secsimulation}
\new{With the geometrical parameters
and mass ratios derived above,
we have constructed N-body/SPH simulations
to demonstrate the formation of the blue ring
and to test the disk survival hypothesis.
We used the N-body/SPH software called ``{\sc Zeno}''
which is developed by
Barnes\footnote{http://www.ifa.hawaii.edu/{$\sim$}barnes/software.html}
\citep[e.g.,][]{1992ARA&A..30..705B}.
Galaxy models are created following
\citet{1988ApJ...331..699B,2002MNRAS.333..481B}.
In the ``{\sc Zeno}'', we have $G\,=\,1$.
The mass ratio between two galaxies is set to be 4:1.
Using the same fraction of halo and baryonic matter as in
\citet{2002MNRAS.333..481B},
the total mass of the system was set to 3.25 mass units.
With the stellar mass of the system,
2.75\,$\times$\,$10^{11}$\,$M_\odot$,
we get the mass unit of simulation as
4.4\,$\times$\,$10^{11}$\,$M_\odot$.
Then we chose the length scale 38.6\,kpc,
resulting in the velocity unit of 221\,km\,s$^{-1}$
and the time unit of 176\,Myrs.
For the main progenitor, we adopted 
B/T=0.34, 
scale length of 5.27\,kpc
and scale height of 2.5\,kpc ;
while for the intruder, B/T=0.65, 
scale length of 3.0\,kpc,
and scale height of 0.5\,kpc.
We assume 5\% of gas distributed in the intruder disk,
and no gas in the main galaxy.
The gas has the same properties as in \citet{2002MNRAS.333..481B},
and the SPH calculation follows the isothermal equation of state.
\begin{figure}[!th]
\centering
\includegraphics[width=8cm]{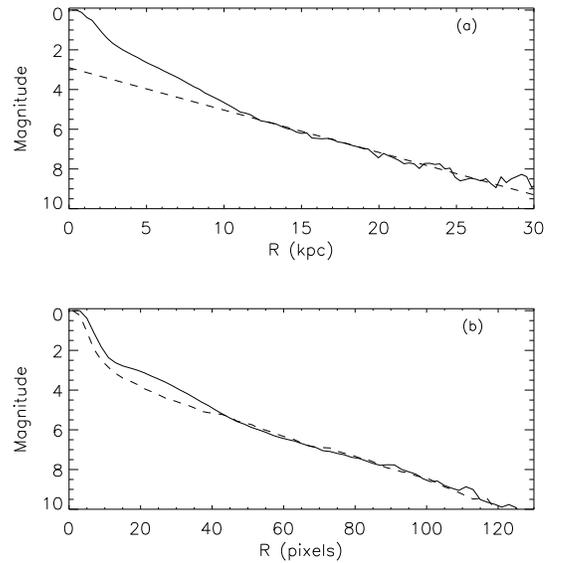}
\caption{Panel ($a$), the observed radial profile (solid line).
The dashed line indicates the slope of the thin-disk component.
At the radii from 5 to 10\,kpc, we clearly have a thick-disk component
which corresponds to the ring that has been formed in the disk of main galaxy.
Panel ($b$), the radial profile derived from the simulation.
Clearly we have a thin- and a thick-disk component.
The appearance of the thick-disk component is related to the formation of the ring.
For comparison, the radial profile of the main progenitor is plotted as dashed line. }
\label{figradialprofiles}
\end{figure}
Our simulations include
$N_{\rm halo}+N_{\rm star}+N_{\rm gas} = 45000 + 44856 + 7560 = 97416$ particles.
The gas fraction in our system is very low.
Therefore, we chose a small mass resolution for SPH calculation
in order to have a reasonable number of particles to trace the gas motion.
We designed a polar collision that follows a parabolic orbit
with pericentric separation of $\sim$\,0.76 kpc.
The fine-tuning of orbital parameters\footnote{
In the {\sc Zeno} system, the exact parameters we adopted for the current model are,
at {\tt t=-0.12} from the first passage,
{\tt deltar=0,0,0.5 deltav=-0.56,0.38,-3.5}, the orientation of intruder
{\tt thetax=35 thetay=-25 thetaz=0} and the orientation of the main
{\tt thetax=18 thetay=18 thetaz=55}
}
is necessary to closely match the morphology and the gas kinematics.
}

\new{
This simulation matches both the morphology and the kinematics that we observed,
as well as some detailed features.
Below, we list those particular features that are matched by the simulation.
The strongest constraints are the ACS morphology and the \GIRAFFE\  kinematics.
\begin{enumerate}
\item[1.]
In Fig.~\ref{figmodel}($d$), we show a simulated image\footnote{
\new{1) All the simulated images in this paper are created 
by the projected density of particles directly. 
Thus the gas component is emphasized by a factor of 50 times due
to the large number of gas particles.
2) All the simulated images, except Fig.~\ref{figrembound}, 
have been scaled to the same physical size as Fig.\ref{figcolors}.}
}
that is constructed by the projected star density
with the gas superposed in blue color. 
The observed blue ring is defined by rest-frame U-band clumps
which indicate the star-burst regions.
Thus the blue ring basically traces the gas.
In the simulation, a gas ring begins to form after the first passage,
and evolves 35\,Myrs to match the observation.
Note that the gas ring in the simulation has a slightly larger size 
than the observed blue ring in order to match the \GIRAFFE\ kinematics.
On the other hand, this is expected since the stellar evolution 
is found to be behind the expanding density wave \citep{1996FCPh...16..111A}.
\item[2.] The kinematics of the object from \GIRAFFE\ is derived
from \oii\ doublet emission that is coming from the ionized gas.
In Fig.~\ref{figmodel}($e$, $f$),
we show kinematics of gas from the simulation.
Geometrically and quantitatively, the simulated gas kinematics is
in good agreement with the observation, i.e.,
the dynamical axis of the velocity field and the peak of the velocity dispersion.
\end{enumerate}
Furthermore, the simulation shows consistencies in several details.
\begin{enumerate}
\item[3.] In Fig.~\ref{figcolors},
  we may notice the slight distortion of the morphology
  at the left part of the ring, close to the edge of the ring.
  This distortion can be seen also in the B$-$z color-map.
  In the simulated image, we find the same feature.
  The distortion is due to the superposition of the two galaxies.
\item[4.] Both the three-color image of Fig.~\ref{figcolors}
  and the residual image of Fig.\ref{majoraxis}
  show a tail-like feature.
  From the simulation we find that it is likely to be
  the tidal tail created by the collision.
  In Fig.~\ref{figfallmain}($a$), we compare the gas
  and the stellar component for the intruder galaxy.
  We notice that the stars move further than the gas,
  marked by the an arrow in the figure.
  Their relative position resembles the blue ring
  and the tail-like structure in the observation.
  The simulation does not show an exact tail structure.
  Nevertheless, the star dynamics has been demonstrated 
  reasonably well.
\begin{figure}[!t]
\centering
\includegraphics[width=9cm]{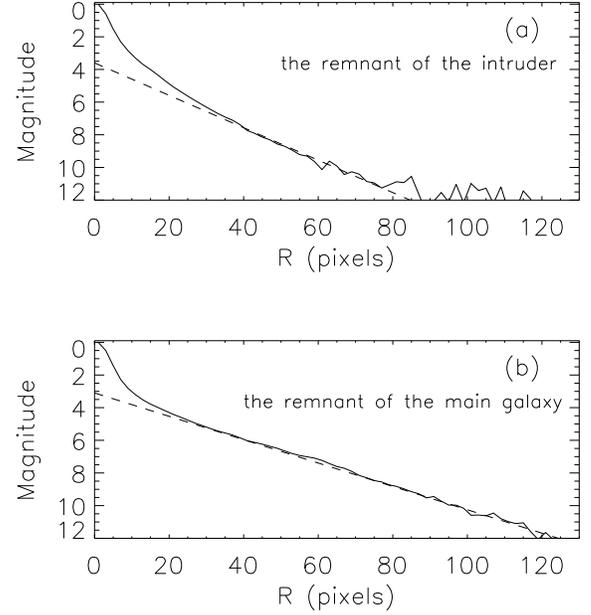}
\caption{The radial profiles of the remnants in the unbound case,
panel~($a$) for the remnant of the intruder,
($b$) for the remnant of the main galaxy.
The dashed lines in both figures indicate the disk components.
Clearly, the remnant of the main progenitor is dominated by the disk.}
\label{figrpremunbound}
\end{figure}
\begin{figure}[ht]
\centering
\includegraphics[width=4.cm]{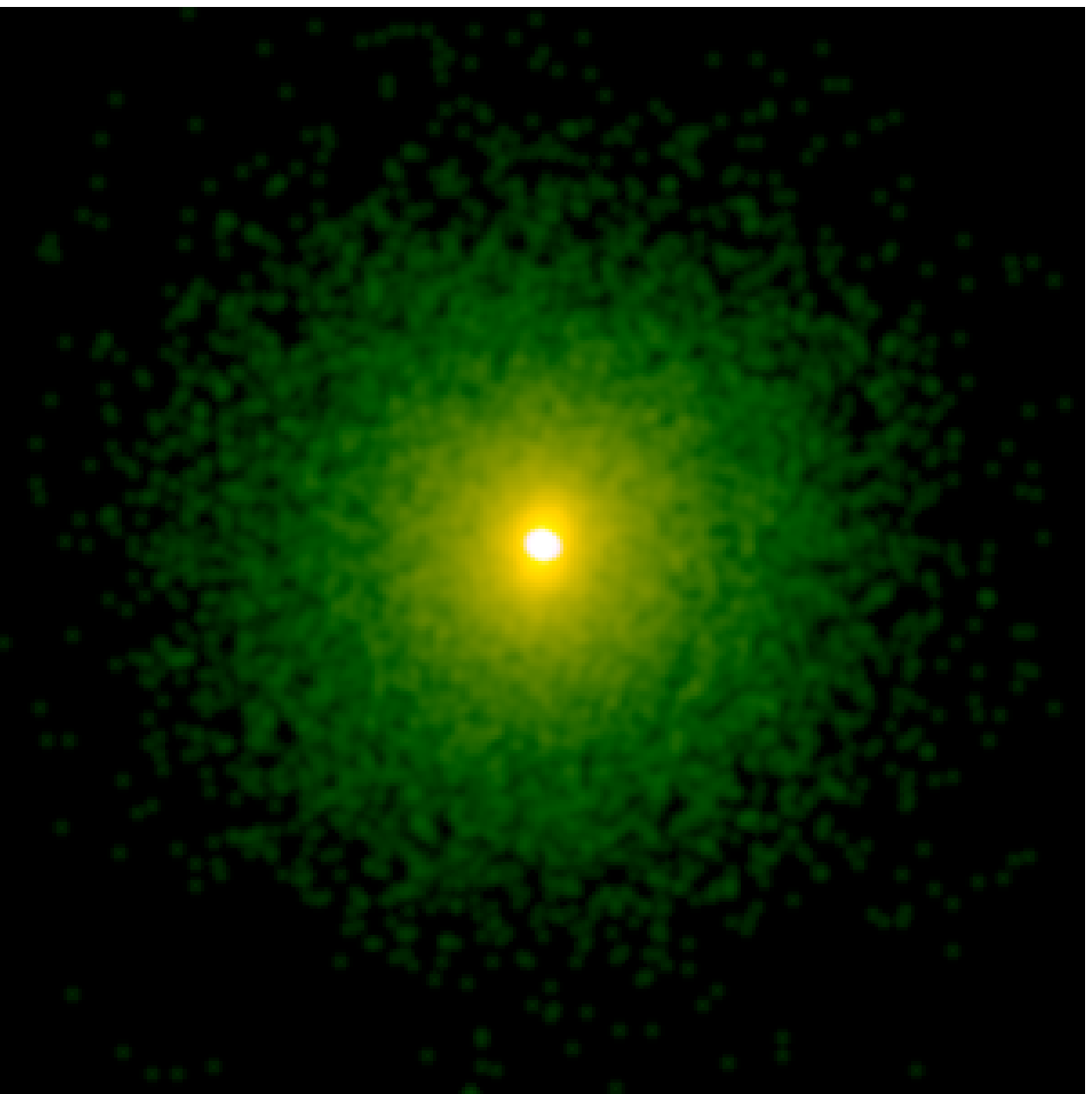}\includegraphics[width=5cm]{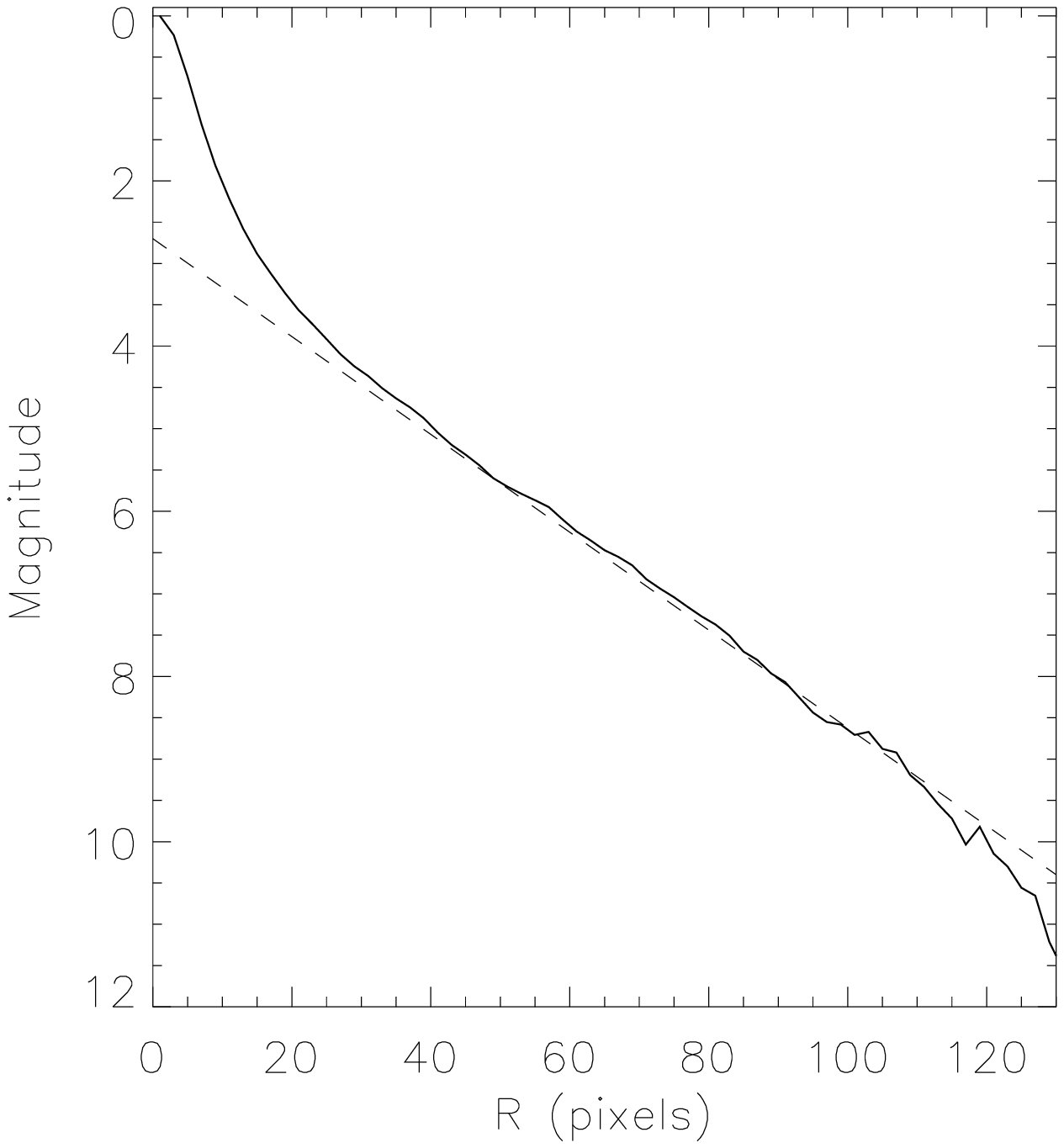}
\caption{{\it Left\/}: The face-on view of the merger remnants. \new{The image has a size of $103$\,kpc}.
{\it Right\/}: the radial profile of the merger remnant. The disk component is indicated by a dashed line.}
\label{figrembound}
\end{figure}
\item[5.] Our target \object{J033210.76--274234.6} is a strong galactic collision.
  We have estimated that the mass ratio of progenitors is not larger than 5\,:\,1,
  and adopted 4:1 in modeling.
  The intruder penetrated the center of the main galaxy.
  A ring in the main galaxy is expected to be observed,
  as we have seen in the simulation (see Fig.~\ref{figfallmain}$b$).
  Although we cannot see the ring directly,
  we have found two robust observational evidences,
  enlightened by the simulation, to prove its existence.
  In Fig.~\ref{figradialprofiles}, we compared the observed radial profile
  with the simulated one. Both of them exhibit two disk components:
  a thick disk and a more extended thin disk.
  Note that for the observation,
  the thin disk is only detected in the radial profile.
  Hence, from the simulation, we may conclude that the thick-disk component
  is caused by the formation of the ring.  This leads to the realization
  that the disk we have seen in the three-color figure is actually the ring
  that has formed in the main galaxy.
  The second evidence is from the residual image (see Fig.~\ref{majoraxis})
  where we have marked a light-deficiency region in the central part.
  It is naturally explained when a disk with a ring is fitted and
  subtracted by a disk profile.
  Therefore, the ring in the main galaxy has been observed.
  However, due to the superposition of the two bright bulges
  and the broadness of the ring, the shape of the ring is hidden.
  The broadness of the ring is linked to the thickness of the main disk,
  which has been investigated and proved by our simulations.
\item[6.] In the previous section, we mentioned that the center of core-B
  in B-band is slightly shifted from other bands,
  resulting the asymmetric color distribution of core-B (see Fig.~\ref{figcolors}).
  In the simulation, we have observed that, after the first passage,
  part of the perturbed gas is falling back to the center of the intruder.
  The falling gas may cause the star-forming activities which could explain
  the color shift of core-B.
  Note that, from the simulation, we find that the maximum velocity
  in the VF map is also caused by the infalling gas.
\item[7.] The morphology decomposition reveals that the bulge of the main progenitor,
  core-A, has been shifted from the center of the disk.
  This is also confirmed by the simulation.
  In Fig.~\ref{figfallmain}$b$, the bulge of the main galaxy
  is clearly shifted with respect to the center of the disk.
  The movement of the central bulge is due to the very close core interaction.
\item[8.] From the FORS2 slit spectrum, 
we have a constraint for the collision speed. 
We measured the velocity difference
between the \oii\ emission and Ca\,{\sc ii}\,H,\,K absorption lines,
finding $210\pm50$\,km\,s$^{-1}$.
Note that by measuring the absorption lines 
we are actually measuring the mean velocity of the two bright bulges 
when considering seeing effects.
Using the simulation, we mimicked the slit observation 
and measured the mean velocity of star particles and of gas particles, respectively.
We find a velocity difference of $270\pm60$\,km\,s$^{-1}$ 
which is consistent with the value from the observation. 
\end{enumerate}
}

\new{
To know if the disk of progenitors will be destroyed by the collision,
we have explored the remnants by simulations.
In our simulation, the collision speed at the first passage is $\sim$1200\,km\,s$^{-1}$  and
the two objects are departing at the speed of $\sim$500\,km\,s$^{-1}$  at the observing time.
They are not a gravitationally bound system.
We keep the simulation running another 2 Gyrs.
The remnants of the both galaxies show a disk component.
Especially, the remnant of the main galaxy is still dominated by disk.
In Fig.~\ref{figrpremunbound}, we plotted the radial profile of the two remnants.
Furthermore, we have run a simulation with lower collision speed,
in order to investigate the remnant
in case that the two galaxies can be merged.
We find that the merger remnant still has a significant disk component
at 3.5 Gyrs after the first passage, see Fig.~\ref{figrembound}.
}

\section{Discussion \& Conclusion}
\label{secdiscuss}
Morpho-kinematics analysis reveals that
\object{J033210.76--274234.6} is dominated by a massive
disk. The spatially-resolved kinematics appears to be 
the superposition of the rotation and the expanding velocity field 
from the blue ring. 
We propose a near-center collision model which
explains all the observations. Consequently, we conclude
that the disk of the main galaxy has survived the galaxy
collision. \new{ Followed by the N-body/SPH simulations,
we have successfully reproduced the ACS morphology
and the \GIRAFFE\ kinematics,
as well as some details.
With the help of the N-body/SPH simulations,
we have explored the remnants of the collision.
We find that this is a rare polar collision
with a very small ($\sim$\,1\,kpc) pericentric separation,
and that regardless of whether the two galaxies merge or not,
the remnant of the main galaxy is dominated by disk component.
The results support the disk survival hypothesis.
The survival of the main disk is related to
the exceptional collision with the extremely small pericentric distance and the polar orbit.
}

\begin{acknowledgements}
We would like to thank the referee for the constructive comments.
We would like to thank Isaura Fuentes-Carrera, Benoit
Neichel and Sebastien Peirani for the helpful discussions.
We are grateful to Albrecht R{\"u}diger for helping us in
the writing of the paper.
Especially, we would like to thank Joshua~E. Barnes for his help on the simulations.
\end{acknowledgements}

\end{document}